# Orbital-Selective Spin Splitting in the Altermagnetic $Ti_2XX'$ Monolayers

*Xin Zhang and Shihao Zhang**

**ABSTRACT:** Altermagnets have attracted intensive attention recently, owing to their unique combination of zero net magnetic moment of antiferromagnets and momentum space spin-splitting of ferromagnets. Herein, we systematically explore monolayer altermagnetic $Ti_2XX'$ (X/X' = F, Cl, Br, I) materials for novel electronic properties. Especifically, spin-splitting exists in the valence bands of all studied materials, whereas several compositions exhibit nearly spin-degenerate conduction bands near the Fermi level, which is markedly different from conventional altermagnets. This phenomenon can be attributed to the bond-angle-sensitive super-exchange occurring in the $d_{xz}$ or $d_{yz}$ orbitals, while the super-exchange of the $d_{x^2-y^2}$ orbital is not sensitive to the Ti-X-Ti bond angle. Motivated by this intriguing phenomenon, Fermi level modulation via doping or heterostructure construction enables the transition between two electronic states, endowing these materials great application potential in information storage devices.

Antiferromagnetic materials have attracted considerable attention in high density, low power spintronic devices owing to their zero net magnetic moment, immunity to external magnetic perturbations, and ultrafast spin dynamics. [1-3] Compared with conventional antiferromagnets, altermagnets combine the zero-magnetic-moment character of antiferromagnets with the momentum space spin-splitting typically associated with ferromagnets. [4,5] Remarkably, altermagnets can give rise to strongly spin-split bands without relying on strong spin-orbit coupling, thereby overcoming the performance limitations of traditional spintronic materials and emerging as a vibrant research frontier in condensed matter physics and materials science. [6-43]

Benefiting from their absence of stray magnetic fields, efficient current-driven manipulation of the Néel vector, and electrical readout of spin signals, altermagnets are regarded as highly promising candidates for next-generation magnetic random-access memories, spin logic devices, and terahertz spintronic devices. Experimentally, altermagnetic signatures have been observed in several material systems, such as MnTe, [44-48] CrSb, [49-52] and $KV_2Se_2O$. [53] Their intrinsic altermagnetic band features and spin-transport properties have been systematically verified and explored, offering promising material platforms for high density spintronic applications. Nevertheless, existing experimental and theoretical studies have mainly focused on conventional altermagnets in which both the valence and conduction bands undergo spin-splitting. The asymmetric altermagnetic state characterized by spin-split valence bands but nearly spin-degenerate conduction bands has rarely been reported, and its underlying microscopic mechanism and tunability remain elusive. Therefore, discovering novel altermagnetic materials with such anomalous band features and elucidating the physical origin of their distinctive spin-transport behavior is of great significance for enriching altermagnetic physics and expanding the family of spintronic materials.

Motivated by this, we systematically investigate the crystal structures, electronic band structures, and spin-transport properties of monolayer $Ti_2XX'$ two-dimensional altermagnets. We find that several $Ti_2XX'$ systems retain the characteristic spin-split valence bands while exhibiting anomalous nearly spin-degenerate conduction bands near the Fermi level, in sharp contrast to conventional altermagnets. Combining orbital-projected band structures, bond-angle analyses, and spin-resolved conductivity simulations, we reveal that the Ti-X-Ti bond angle governs the super-exchange interaction of the $d_{xz}$ and $d_{yz}$ orbitals, which is responsible for the emergence of conduction-band spin-degeneracy. Owing to this unique asymmetric spin-splitting feature, the Fermi level position in these materials can be precisely tuned via element doping or charge transfer induced by heterostructure construction, enabling controllable switching between spin-degenerate and spin-split electronic states. This work not only identifies a class of two-dimensional altermagnetic materials with exotic band characteristics but also provides new insights and theoretical guidance for designing high-performance, tunable spintronic devices for information storage.

Our first-principles calculations were conducted within the framework of density functional theory (DFT) by employing the DS-PAW module integrated in the Device Studio platform with the projector-augmented wave (PAW) pseudopotential scheme.

[54,55] The Perdew–Burke–Ernzerhof (PBE) functional was used to treat electron exchange-correlation interactions. For geometric optimization and self-consistent calculations, the convergence criteria were fixed at $10^{-6}$ eV for total energy and 0.05 eV/Å for atomic forces, with a plane-wave kinetic-energy cutoff of 600 eV applied throughout all computations. Moreover, a vacuum layer of more than 20 Å was constructed to suppress artificial interlayer coupling between periodically repeated images. The Brillouin zone was sampled via a Γ-centered 7 × 7 × 1 Monkhorst-Pack *k*-point mesh. [56] To account for on-site Coulomb interaction, DFT+$U$ calculations were performed with an effective Hubbard parameter $U_{\mathrm{eff}} = U - J = 3.0$ eV assigned to the *d*-orbitals of Ti atoms. [57,58] We also performed the electronic structure calculations with the HSE06 screened hybrid functional using the VASP package. [59-61]

Monolayer $Ti_2XX'$ adopts a square-lattice structure with an X'-Ti-X atomic sequence, forming a sandwich-type geometry, as shown in Figure 1(a, b). $Ti_2XX'$ belongs to the *P*4*mm* space group (No. 99). When X = X', the system yields the $Ti_2X_2$ structure, which falls into the *P*4/*mmm* space group (No. 123) and possesses an additional $\mathcal{M}_z$ symmetry. Our calculations reveal that such structural difference is irrelevant to the properties observed in this work. Ti atoms adopt an A-type antiferromagnetic configuration, with magnetic moments aligned along the + *z* and - *z* directions, as shown in Figure 1(a). The spin-splitting behavior observed in the band structures (Figure 2, S1 and S2) confirms the altermagnetic nature of these materials. Notably, $Ti_2Cl_2$, $Ti_2Br_2$, $Ti_2ClBr$, $Ti_2ClI$ and $Ti_2BrI$ possess nearly spin-degenerate conduction bands near the Fermi level. This feature differs from $Ti_2F_2$, $Ti_2FCl$, $Ti_2FBr$ and $Ti_2FI$, whose conduction bands also exhibit spin-splitting, consistent with the behavior of conventional altermagnets. Figure 1(c) shows a schematic of spin-split valence bands and nearly spin-degenerate conduction bands. The separated blue and red curves in the valence band represent the spin-up and spin-down components, demonstrating prominent spin-splitting. By contrast, the purple curve in the conduction band denotes the overlapping blue and red traces, indicative of spin-degeneracy. Herein, we take $Ti_2Br_2$ as a representative example to analyze this peculiar phenomenon.

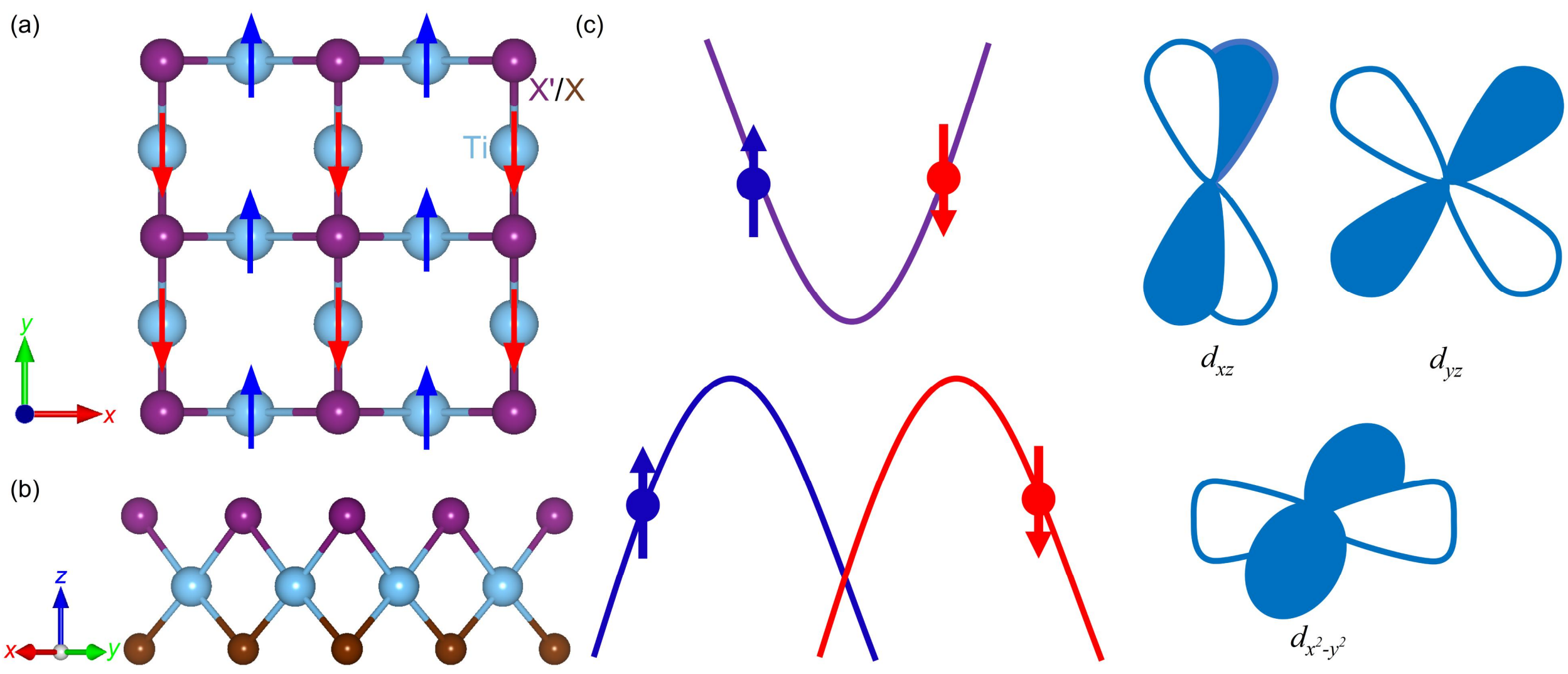

**Figure 1.** (a) Top view and (b) side view of schematic structures of $Ti_2XX'$. Light-blue, brown and purple spheres correspond to Ti, X and X' atoms, respectively. Blue and red arrows mark the atomic magnetic moments of the antiferromagnetic configuration aligned along the + *z* and - *z* directions, respectively. (c) Schematic of spin-split valence bands and spin-degenerate conduction bands. Blue and red curves represent spin-up and spin-down bands, while purple curves denote nearly spin-degenerate bands. The conduction band originates primarily from $d_{xz}$ and $d_{yz}$ orbitals, whereas the valence band is mainly derived from the $d_{x^2-y^2}$ orbital.

Figure 2 presents the calculated band- structure results of $Ti_2Br_2$. Figure 2(a) shows the results obtained with the PBE+$U$ functional ($U$ = 3 eV), where the material exhibits metallic behavior with no band gap. Obvious splitting between spin- up (blue curves) and spin- down (red curves) channels can be observed in the valence bands, which is a hallmark of altermagnets. Nevertheless, the spin- up and spin- down channels overlap in the conduction bands, maintaining nearly spin-degeneracy over the energy range of 0~1.35 eV. This behavior deviates from conventional altermagnets whose conduction bands also feature spin-splitting, indicating the emergence of a novel physical property.

To achieve a more reliable description of spin- splitting and band- gap values beyond the PBE- GGA level, we employed the HSE06 screened hybrid functional. The corresponding band- structure results are displayed in Figure 2(b). Although the material becomes a semiconductor with a band gap of approximately 0.8 eV near the Fermi level, which arises from the well- known tendency of HSE06 functional to produce larger band gaps compared to PBE- GGA. The band structure still retains spin- split valence bands and nearly spin- degenerate conduction bands, further confirming this peculiar property.

To further explore the origin of this behavior, we performed projected band- structure calculations, as illustrated in Figure 2(c, d). Figure 2(c) and figure 2(d) present the projected band weight curves of Ti $d$ orbitals for spin-up and spin-down channels, respectively. The valence bands near the Fermi level are predominantly contributed by the $d_{x^2-y^2}$ orbitals and exhibit prominent spin-splitting. In contrast, the conduction bands around the Fermi level originating from $d_{xz}$ and $d_{yz}$ orbitals maintain spin-degeneracy, which is likely attributed to the weak super-exchange interaction associated with these two orbitals. To further clarify the underlying mechanism, we systematically analyzed the Ti-X-Ti and Ti-X'-Ti bond angles, as summarized in Figure S3(a, b). The results reveal that systems with nearly spin-degenerate conduction bands possess smaller bond angles below 68°, whereas conventional altermagnets with spin-split conduction bands exhibit larger bond angles exceeding 74°. This distinct angular difference further verifies that reduced bond angles weaken the super-exchange effect of $d_{xz}$ and $d_{yz}$ orbitals, thereby inducing conduction band spin-degeneracy.

Within the ligand-mediated super-exchange picture, the effective $d$-electron transfer between neighboring Ti sites arises from a second-order virtual process through the halogen $p$ states, $t_{eff}(\theta) = {t_{pd}}^2(\theta)/\Delta$, where $\Delta = \varepsilon_d - \varepsilon_p$ is the charge-transfer energy that controls the ligand-to-metal charge-transfer gap in the Zaanen-Sawatzky-Allen classification, [62] and $\varepsilon_p$ and $\varepsilon_d$ denote the halogen $p$ and Ti $d$ orbital energies, respectively. In the Slater–Koster two-center approximation, the hybridization between the Ti $d_{xz}/d_{yz}$ orbitals and the halogen $p$ orbitals scales with the direction cosines of the Ti–X bond, namely $t_{pd}(\theta) \propto \sin(\frac{\theta}{2})\cos(\frac{\theta}{2})$. Combining these factors with Anderson's kinetic-exchange relation $J = 2{t_{eff}}^2/U$ gives

$$J(\theta) \propto {t_{eff}}^2(\theta)/U \propto \sin^4(\frac{\theta}{2})\cos^4(\frac{\theta}{2})(\frac{d_0}{d})^{4n}$$

where $U$ denotes the effective on-site Coulomb interaction and $n \approx 2$ is the conventional Harrison scaling exponent for transfer integrals, with $d_0$ the reference Ti–X bond length.

For a fixed Ti–Ti distance, reducing the bond angle from ≈ 74° to ≈ 68° elongates the Ti–X bond by ≈ 8%, which, together with the angular factor, suppresses *J(θ)* by nearly half of magnitude. The super-exchange interaction of the $d_{xz}$ and $d_{yz}$ orbitals is thus effectively switched off in the small-$\theta$ systems, which removes the exchange field acting on the corresponding conduction states and preserves their spin degeneracy. By contrast, the large-$\theta$ systems recover the conventional $d$-wave altermagnetic splitting of the conduction bands. The $d_{x^2-y^2}$ spin splitting of the orbitals is not sensitive to the Ti-X-Ti and Ti-X'-Ti bond angles, which gives rise to the robust spin splitting of valence bands in different $Ti_2XX'$ monolayers.

The $\Delta E$ values displayed on the right $y$-axis quantify the spin-splitting magnitude at the X point in momentum space, as shown in Figure S3(c). Although spin-degenerate systems still exhibit finite splitting energies, such observable splitting occurs above 1.3 eV relative to the Fermi level and is far away from the Fermi surface. In comparison, conventional altermagnets show spin-splitting at energies as low as 0.4 eV above the Fermi level, which is much closer to the Fermi surface. Accordingly, the exotic band feature observed in our studied systems remains physically valid and fundamentally distinct from traditional altermagnetic behavior.

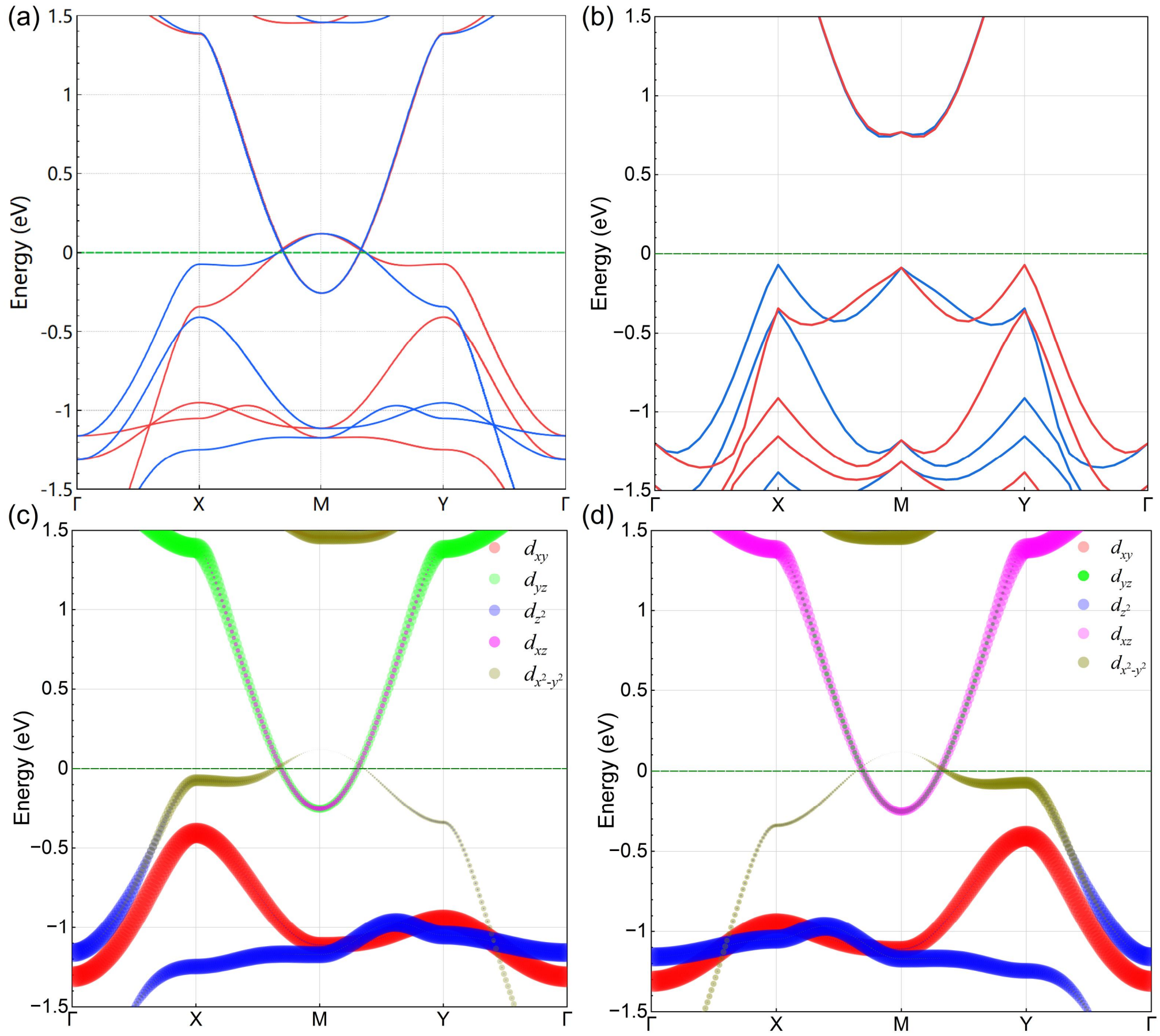

**Figure 2.** Band structures of $Ti_2Br_2$ calculated by different functionals: (a) PBE+$U$ functional with $U$ = 3 eV, (b) HSE06 hybrid functional, (c) projected band weight for spin-up Ti-$d$ orbitals, (d) projected band weight for spin-down Ti-$d$ orbitals.

A minimal effective tight-binding model is adopted to describe the electronic structures of the investigated monolayer system. The Hamiltonian of the monolayer reads: [63]

$$H_0 = [\mu + A(\cos k_x + \cos k_y)]\tau_0\sigma_0 + (\cos k_x - \cos k_y)(B\tau_z\sigma_0 + C\tau_0\sigma_z)$$

$$+ t\cos\frac{k_x}{2}\cos\frac{k_y}{2}\tau_x\sigma_0 + [u + D(\cos k_x + \cos k_y)]\tau_z\sigma_z\,.$$

Here $\tau$ and $\sigma$ denote the Pauli matrices defined in the sublattice and spin space, respectively. The system preserves mirror crystalline symmetry and $C_{4z}\mathcal{T}$ symmetry. For simplicity, spin-orbit coupling is neglected in this work. This microscopic Hamiltonian well reproduces the characteristic band features of altermagnetic systems, as shown in Fig. 3(a). The valence bands exhibit evident spin-splitting, whereas the conduction bands stay spin-degenerate.

To further interpret this feature, we simulated the spin-resolved conductivity $\sigma_{xx}^{\sigma}$ along the $x$ direction under $x$-direction electric field according to the formula given below,

$$\sigma_{xx}^{\sigma}(\mu) = \frac{2\pi}{\Omega}\sum_{\mathbf{k}} Tr\,[v_x^{\sigma} A_{\sigma} v_x^{\sigma} A_{\sigma}]$$

Here $v_x^{\sigma}$ is the velocity operator along $x$ direction of $\sigma$-channel Hamiltonian, and $A_{\sigma}(\mathbf{k},\mu) = -\frac{1}{\pi} Im\, G_{\sigma}^{R}(\mathbf{k},\mu)$ is responding spectrum function where $G_{\sigma}^{\mathrm{R}}$ is retarded Green function. The simulated results are presented in Figure 3(b). One can clearly observe that the spin-up and spin-down conductivities are separated in the valence-band region while becoming degenerate within the conduction-band region, which is consistent with our band-structure simulations. The total conductivity is given by $\sigma_{xx\,tot} = \sigma_{xx}^{\uparrow} + \sigma_{xx}^{\downarrow}$, which is represented by the black curve. In our altermagnetic system, $\sigma_{xx}^{\uparrow} = \sigma_{yy}^{\downarrow}$ and $\sigma_{xx}^{\downarrow} = \sigma_{yy}^{\uparrow}$ due to $C_{4z}\mathcal{T}$ symmetry. Figure 3(c) shows the curve of spin-polarized conductivity, where $P = \frac{\sigma_{xx}^{\uparrow} - \sigma_{xx}^{\downarrow}}{\sigma_{xx}^{\uparrow} + \sigma_{xx}^{\downarrow}}$. [64] This quantity adopts non-zero values in the valence-band region, consistent with unequal spin-up and spin-down conductivities originating from spin-splitting. By contrast, it drops to zero in the conduction-band region, which corresponds to identical spin-resolved conductivities and thus reflects spin-degeneracy.

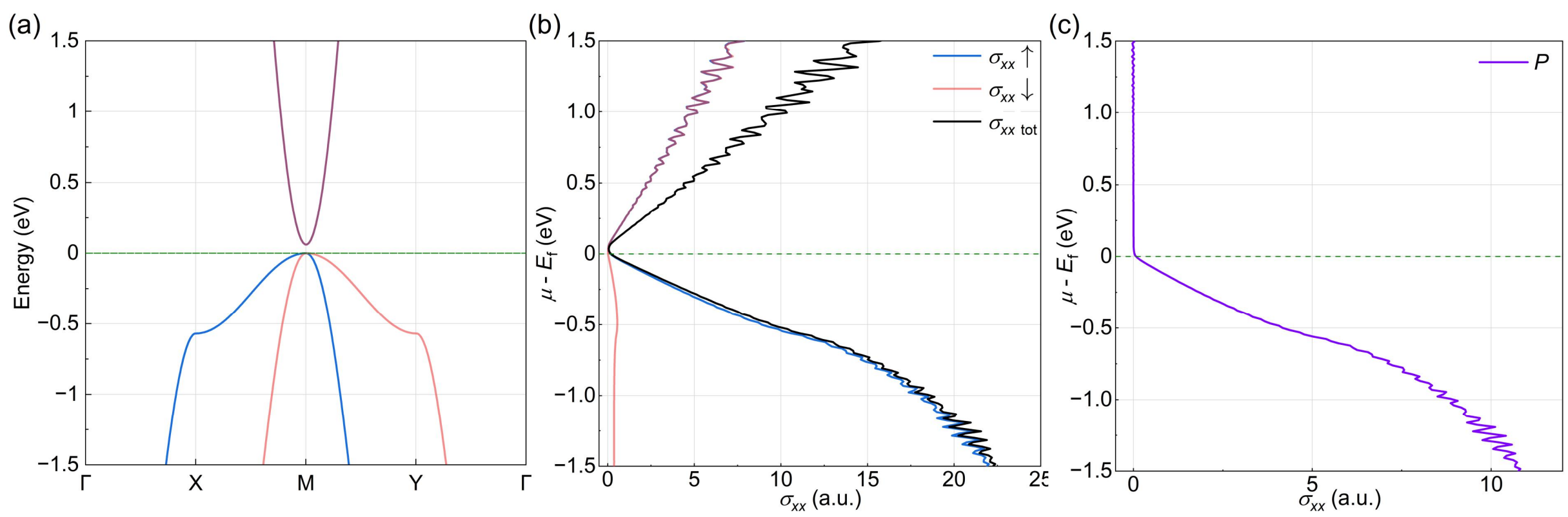

**Figure 3.** (a) Theoretically simulated band structures. We use the set of parameters: $\mu = 0.0,\ A = 1.2, B =- C = u = 0.5, D = 2.485,$ and $t = 3.0$. (b) Spin-resolved conductivity $\sigma_{xx}$ near the Fermi level. The blue, red, and black curves represent spin-up conductivity, spin-down conductivity, and total conductivity, respectively, while the purple curve denotes spin-degenerate conductivity. (c) Spin polarization ($P = \frac{\sigma_{xx}^{\uparrow} - \sigma_{xx}^{\downarrow}}{\sigma_{xx}^{\uparrow} + \sigma_{xx}^{\downarrow}}$ ) near the Fermi level.

In summary, we have investigated the band-structure properties of monolayer altermagnetic $Ti_2XX'$ materials. We find that $Ti_2Cl_2$, $Ti_2Br_2$, $Ti_2ClBr$, $Ti_2ClI$ and $Ti_2BrI$ exhibit nearly spin-degenerate conduction bands near the Fermi level, which deviates from the behavior of conventional altermagnets. Further analysis of Ti-*d* orbital projected band densities suggests that small Ti-X-Ti bond angles weaken the super-exchange interaction of $d_{xz}$ and $d_{yz}$ orbitals, giving rise to such spin-degeneracy. Combined analyses of band structures and spin-resolved conductivities reveal the physical origin behind this peculiar feature. This exotic property enables the modulation of the Fermi-level position via element doping or heterostructure construction to achieve switching between two distinct electronic states. Therefore, these materials hold great promise for device applications toward improved efficiency of information storage.

## ASSOCIATED CONTENT

### Supporting Information

The Supporting Information is available free of charge.

Band structures of $Ti_2Cl_2$, $Ti_2ClBr$, $Ti_2ClI$, $Ti_2BrI$, $Ti_2F_2$, $Ti_2FCl$, $Ti_2FBr$ and $Ti_2FI$ calculated using the PBE + $U$ functional. Bond angles for Ti-X-Ti and Ti-X'-Ti configurations and band-splitting magnitudes at the X point in momentum space of diverse materials (PDF)

## AUTHOR INFORMATION

### Corresponding Author

**Shihao Zhang** - *School of Physics and Electronics, Hunan University, Changsha 410082, China*; Email: zhangshh@hnu.edu.cn

### Author

**Xin Zhang** - *School of Physical Science and Technology, Kunming University, Kunming 650214, China*

### Notes

The authors declare no competing financial interest.

**ACKNOWLEDGMENT**

This work is supported by the National Natural Science Foundation of China (Grant No. 12304217), the National Key Research and Development Program of China (Grant No. 2024YFA1410300), the Natural Science Foundation of Hunan Province (Grant No. 2025JJ60002), the Fundamental Research Funds for the Central Universities from China (Grant No. 531119200247), the Special Basic Cooperative Research Programs of Yunnan Provincial Undergraduate Universities' Association (Grant No. 202401BA070001-013), and Xing Dingyu Academician Workstation of Yunnan Province (202605AF350035). We gratefully acknowledge HZWTECH for providing computation facilities.

**REFERENCES**

(1) Baltz, V.; Manchon, A.; Tsoi, M.; Moriyama, T.; Ono, T.; Tserkovnyak, Y. Antiferromagnetic spintronics. *Rev. Mod. Phys.* **2018**, *90* (1), 015005.
(2) Jungwirth, T.; Marti, X.; Wadley, P.; Wunderlich, J. Antiferromagnetic spintronics. *Nat. Nanotechnol.* **2016**, *11* (3), 231-241.
(3) Šmejkal, L.; Marmodoro, A.; Ahn, K.-H.; González-Hernández, R.; Turek, I.; Mankovsky, S.; Ebert, H.; D'Souza, S. W.; Šipr, O.; Sinova, J.; Jungwirth, T. Chiral Magnons in Altermagnetic $RuO_2$. *Phys. Rev. Lett.* **2023**, *131* (25), 256703.
(4) Šmejkal, L.; Sinova, J.; Jungwirth, T. Beyond Conventional Ferromagnetism and Antiferromagnetism: A Phase with Nonrelativistic Spin and Crystal Rotation Symmetry. *Phys. Rev. X* **2022**, *12* (3), 031042.
(5) Šmejkal, L.; Sinova, J.; Jungwirth, T. Emerging Research Landscape of Altermagnetism. *Phys. Rev. X* **2022**, *12* (4), 040501.
(6) Tschirner, T.; Keßler, P.; Gonzalez Betancourt, R. D.; Kotte, T.; Kriegner, D.; Büchner, B.; Dufouleur, J.; Kamp, M.; Jovic, V.; Smejkal, L.; Sinova, J.; Claessen, R.; Jungwirth, T.; Moser, S.; Reichlova, H.; Veyrat, L. Saturation of the anomalous Hall effect at high magnetic fields in altermagnetic $RuO_2$. *APL Mater.* **2023**, *11* (10).
(7) Bai, L.; Feng, W.; Liu, S.; Šmejkal, L.; Mokrousov, Y.; Yao, Y. Altermagnetism: Exploring New Frontiers in Magnetism and Spintronics. *Adv. Funct. Mater.* **2024**, *34* (49), 2409327.
(8) Guo, S.-D.; Liu, Y.; Yu, J.; Liu, C.-C. Valley polarization in twisted altermagnetism. *Phys. Rev. B* **2024**, *110* (22), L220402.
(9) Krempaský, J.; Šmejkal, L.; D'Souza, S. W.; Hajlaoui, M.; Springholz, G.; Uhlířová, K.; Alarab, F.; Constantinou, P. C.; Strocov, V.; Usanov, D.; Pudelko, W. R.; González-Hernández, R.; Birk Hellenes, A.; Jansa, Z.; Reichlová, H.; Šobáň, Z.; Gonzalez Betancourt, R. D.; Wadley, P.; Sinova, J.; Kriegner, D.; Minár, J.; Dil, J. H.; Jungwirth, T. Altermagnetic lifting of Kramers spin degeneracy. *Nature* **2024**, *626* (7999), 517-522.
(10) Leiviskä, M.; Rial, J.; Bad'ura, A.; Seeger, R. L.; Kounta, I.; Beckert, S.; Kriegner, D.; Joumard, I.; Schmoranzerová, E.; Sinova, J.; Gomonay, O.; Thomas, A.; Goennenwein, S. T. B.; Reichlová, H.; Šmejkal, L.; Michez, L.; Jungwirth, T.; Baltz, V. Anisotropy of the anomalous Hall effect in thin films of the altermagnet candidate $Mn_5Si_3$. *Phys. Rev. B* **2024**, *109* (22), 224430.
(11) Qi, Y.; Zhao, J.; Zeng, H. Spin-layer coupling in two-dimensional altermagnetic bilayers with tunable spin and valley splitting properties. *Phys. Rev. B* **2024**, *110* (1), 014442.
(12) Rao, P.; Mook, A.; Knolle, J. Tunable band topology and optical conductivity in altermagnets. *Phys. Rev. B* **2024**, *110* (2), 024425.
(13) Reichlova, H.; Lopes Seeger, R.; González-Hernández, R.; Kounta, I.; Schlitz, R.; Kriegner, D.; Ritzinger, P.; Lammel, M.; Leiviskä, M.; Birk Hellenes, A.; Olejník, K.; Petřiček, V.; Doležal, P.; Horak, L.; Schmoranzerova, E.; Badura, A.; Bertaina, S.; Thomas, A.; Baltz, V.; Michez, L.; Sinova, J.; Goennenwein, S. T. B.; Jungwirth, T.; Šmejkal, L. Observation of a spontaneous anomalous Hall response in the $Mn_5Si_3$ d-wave altermagnet candidate. *Nat. Commun.* **2024**, *15* (1), 4961.

(14) Wu, Y.; Deng, L.; Yin, X.; Tong, J.; Tian, F.; Zhang, X. Valley-Related Multipiezo Effect and Noncollinear Spin Current in an Altermagnet $Fe_2Se_2O$ Monolayer. *Nano Lett.* **2024**, *24* (34), 10534-10539.
(15) Zhou, X.; Feng, W.; Zhang, R.-W.; Šmejkal, L.; Sinova, J.; Mokrousov, Y.; Yao, Y. Crystal Thermal Transport in Altermagnetic $RuO_2$. *Phys. Rev. Lett.* **2024**, *132* (5), 056701.
(16) Zhu, Y.; Chen, T.; Li, Y.; Qiao, L.; Ma, X.; Liu, C.; Hu, T.; Gao, H.; Ren, W. Multipiezo Effect in Altermagnetic $V_2SeTeO$ Monolayer. *Nano Lett.* **2024**, *24* (1), 472-478.
(17) Zhu, Y.-P.; Chen, X.; Liu, X.-R.; Liu, Y.; Liu, P.; Zha, H.; Qu, G.; Hong, C.; Li, J.; Jiang, Z.; Ma, X.-M.; Hao, Y.-J.; Zhu, M.-Y.; Liu, W.; Zeng, M.; Jayaram, S.; Lenger, M.; Ding, J.; Mo, S.; Tanaka, K.; Arita, M.; Liu, Z.; Ye, M.; Shen, D.; Wrachtrup, J.; Huang, Y.; He, R.-H.; Qiao, S.; Liu, Q.; Liu, C. Observation of plaid-like spin splitting in a noncoplanar antiferromagnet. *Nature* **2024**, *626* (7999), 523-528.
(18) Chang, Y.; Wu, Y.; Deng, L.; Yin, X.; Zhang, X. Valley-Related Multipiezo Effect in Altermagnet Monolayer $V_2STeO$. *Materials* **2025**, *18* (3), 527.
(19) Duan, X.; Zhang, J.; Zhu, Z.; Liu, Y.; Zhang, Z.; Žutić, I.; Zhou, T. Antiferroelectric Altermagnets: Antiferroelectricity Alters Magnets. *Phys. Rev. Lett.* **2025**, *134* (10), 106801.
(20) Gu, M.; Liu, Y.; Zhu, H.; Yananose, K.; Chen, X.; Hu, Y.; Stroppa, A.; Liu, Q. Ferroelectric Switchable Altermagnetism. *Phys. Rev. Lett.* **2025**, *134* (10), 106802.
(21) He, X.; Zhang, S. Dirac fermions in the altermagnet $Ce_4Sb_3$. *Phys. Rev. B* **2025**, *112* (7), 075138.
(22) Hu, X.; Zhao, W.; Xia, W.; Sun, H.; Wu, C.; Wu, Y.-Z.; Li, P. Valley polarization and anomalous valley Hall effect in altermagnet $Ti_2Se_2S$ with multipiezo properties. *Appl. Phys. Lett.* **2025**, *127* (1).
(23) Lei, B.; Li, A.; Duan, H.; Long, M.; Wang, Y.; Ouyang, F. Shear-strain-induced switchable spin splitting and piezomagnetic properties in altermagnetic materials. *Front. Phys.* **2025**, *20* (6), 064204.
(24) Li, Y.-Q.; Zhang, Y.-K.; Lu, X.-L.; Shao, Y.-P.; Bao, Z.-Q.; Zheng, J.-D.; Tong, W.-Y.; Duan, C.-G. Ferrovalley Physics in Stacked Bilayer Altermagnetic Systems. *Nano Lett.* **2025**, *25* (15), 6032-6039.
(25) Qu, S.; Hou, X.-Y.; Liu, Z.-X.; Guo, P.-J.; Lu, Z.-Y. Altermagnetic Weyl node-network metals protected by spin symmetry. *Phys. Rev. B* **2025**, *111* (19), 195138.
(26) Sun, H.; Dong, P.; Wu, C.; Li, P. Multifield-induced antiferromagnet transformation into altermagnet and realized anomalous valley Hall effect in monolayer $VPSe_3$. *Phys. Rev. B* **2025**, *111* (23), 235431.
(27) Vila, M.; Sunko, V.; Moore, J. E. Orbital-spin locking and its optical signatures in altermagnets. *Phys. Rev. B* **2025**, *112* (2), L020401.
(28) Wang, D.; Wang, H.; Liu, L.; Zhang, J.; Zhang, H. Electric-Field-Induced Switchable Two-Dimensional Altermagnets. *Nano Lett.* **2025**, *25* (1), 498-503.
(29) Yi, X.-J.; Mao, Y.; Lu, X.; Sun, Q.-F. Spin splitting Nernst effect in altermagnets. *Phys. Rev. B* **2025**, *111* (3), 035423.

(30) Zeng, H.; Zhang, W.; Qiu, C.; Ding, D.-Z.; Zhao, J. Symmetry breaking induced nonrelativistic spin splitting and spontaneous valley polarization in altermagnetic $Ca(CoN)_2$ bilayer. *Appl. Phys. Lett.* **2025**, *126* (20).
(31) Zhou, J.; Zhang, C. Contrasting Light-Induced Spin Torque in Antiferromagnetic and Altermagnetic Systems. *Phys. Rev. Lett.* **2025**, *134* (17), 176902.
(32) Zou, X.; Feng, X.; Dai, Y.; Huang, B.; Niu, C. Floquet Quantum Anomalous Hall Effect with In-Plane Magnetization in Two-Dimensional Altermagnets. *ACS Nano* **2025**, *19* (40), 35575-35580.
(33) Guo, S.-D. Hidden altermagnetism. *Front. Phys.* **2026**, *21* (2), 025201.
(34) Guo, S.-D.; Luo, Q.; Zhang, S.-H.; Jiang, P. External field induced transition from altermagnetic metal to fully compensated ferrimagnetic metal in monolayer $Cr_2O$. *Phys. Rev. B* **2026**, *113* (6), 064408.
(35) Qu, S.; Ouyang, Z.-F.; Gao, Z.-F.; Sun, H.; Liu, K.; Guo, P.-J.; Lu, Z.-Y. Extremely strong spin−orbit coupling effect in light-element altermagnetic materials. *Front. Phys.* **2026**, *21* (4), 045203.
(36) Tan, C.-Y.; Gao, Z.-F.; Yang, H.-C.; Liu, K.; Guo, P.-J.; Lu, Z.-Y. Bipolarized Weyl semimetals and quantum crystal valley Hall effect in two-dimensional altermagnetic materials. *Phys. Rev. B* **2026**, *114* (5), 055125.
(37) Xu, R.; Gao, Y.; Liu, J. Chemical design of monolayer altermagnets. *Natl. Sci. Rev.* **2026**, *13* (2), nwaf528.
(38) Zhang, R.-W.; Cui, C.; Wang, Y.; Duan, J.; Yu, Z.-M.; Yao, Y. Quantized spin Hall conductivity in altermagnetic $Fe_2Te_2O$ with mirror-spin coupling. *Phys. Rev. B* **2026**, *113* (16), L161115.
(39) Zhang, X.; Zhang, S. Sliding-induced ferrovalley polarization and possible antiferromagnetic half-metal in bilayer altermagnets. *Front. Phys.* **2026**, *21* (7), 075203.
(40) Zhang, S. Third-order optical response in *d*-wave altermagnets: Analytical and numerical results from a microscopic model. *Phys. Rev. B* **2026**.
(41) Zhang, S. Intrinsic antiferromagnetic half-metal and topological phases from the ferrovalley states of the sliding bilayer altermagnets. *npj Quantum Mater.* **2026**.
(42) Zhang, T.; Chen, X.; Zhang, S. Altermagnetism in $CoNb_4Se_8$ as Pomeranchuk Instability. *Chin. Phys. B* **2026**.
(43) Zhang, S. Quantum-metric-driven light-induced ferrovalley state in d-wave altermagnets. **2026**, arXiv:2607.17049.
(44) Gray, I.; Deng, Q.; Tian, Q.; Chilcote, M.; Dodge, J. S.; Brahlek, M.; Wu, L. Time-resolved magneto-optical effects in the altermagnet candidate MnTe. *Appl. Phys. Lett.* **2024**, *125* (21).
(45) Liu, Z.; Ozeki, M.; Asai, S.; Itoh, S.; Masuda, T. Chiral Split Magnon in Altermagnetic MnTe. *Phys. Rev. Lett.* **2024**, *133* (15), 156702.
(46) Osumi, T.; Souma, S.; Aoyama, T.; Yamauchi, K.; Honma, A.; Nakayama, K.; Takahashi, T.; Ohgushi, K.; Sato, T. Observation of a giant band splitting in altermagnetic MnTe. *Phys. Rev. B* **2024**, *109* (11), 115102.
(47) Yamamoto, R.; Turnbull, L. A.; Schmidt, M.; Corsaletti Filho, J. C.; Binger, H. J.; Di Pietro Martínez, M.; Weigand, M.; Finizio, S.; Prots, Y.; Ferguson, G. M.; Vool, U.;

Wintz, S.; Donnelly, C. Altermagnetic nanotextures revealed in bulk MnTe. *Phys. Rev. Appl.* **2025**, *24* (3), 034037.
(48) Yang, X.; Cheng, X.; Deng, Z.; Gao, Y.-H.; Yang, Q.-L.; Chang, Z.; Yang, P.-T.; Feng, H.-M.; Zhang, X.-Q.; He, W.; Liu, J.; Cheng, Z.-H. Ultrafast Magneto-Optical Fingerprints of Altermagnetism in MnTe. *Phys. Rev. Lett.* **2026**, *136* (25), 256702.
(49) Ding, J.; Jiang, Z.; Chen, X.; Tao, Z.; Liu, Z.; Li, T.; Liu, J.; Sun, J.; Cheng, J.; Liu, J.; Yang, Y.; Zhang, R.; Deng, L.; Jing, W.; Huang, Y.; Shi, Y.; Ye, M.; Qiao, S.; Wang, Y.; Guo, Y.; Feng, D.; Shen, D. Large Band Splitting in g-Wave Altermagnet CrSb. *Phys. Rev. Lett.* **2024**, *133* (20), 206401.
(50) Bai, Y.; Xiang, X.; Pan, S.; Zhang, S.; Chen, H.; Chen, X.; Han, Z.; Xu, G.; Xu, F. Nonlinear field dependence of Hall effect and high-mobility multi-carrier transport in an altermagnet CrSb. *Appl. Phys. Lett.* **2025**, *126* (4).
(51) Zhou, Z.; Cheng, X.; Hu, M.; Chu, R.; Bai, H.; Han, L.; Liu, J.; Pan, F.; Song, C. Manipulation of the altermagnetic order in CrSb via crystal symmetry. *Nature* **2025**, *638* (8051), 645-650.
(52) Reimers, S.; Odenbreit, L.; Šmejkal, L.; Strocov, V. N.; Constantinou, P.; Hellenes, A. B.; Jaeschke Ubiergo, R.; Campos, W. H.; Bharadwaj, V. K.; Chakraborty, A.; Denneulin, T.; Shi, W.; Dunin-Borkowski, R. E.; Das, S.; Kläui, M.; Sinova, J.; Jourdan, M. Direct observation of altermagnetic band splitting in CrSb thin films. *Nat. Commun.* **2024**, *15* (1), 2116.
(53) Jiang, B.; Hu, M.; Bai, J.; Song, Z.; Mu, C.; Qu, G.; Li, W.; Zhu, W.; Pi, H.; Wei, Z.; Sun, Y.-J.; Huang, Y.; Zheng, X.; Peng, Y.; He, L.; Li, S.; Luo, J.; Li, Z.; Chen, G.; Li, H.; Weng, H.; Qian, T. A metallic room-temperature d-wave altermagnet. *Nat. Phys.* **2025**, *21* (5), 754-759.
(54) Blöchl, P. E. Projector augmented-wave method. *Phys. Rev. B* **1994**, *50* (24), 17953-17979.
(55) Hongzhiwei Technology, Device Studio, Version V2024A, China. **2024**, Avaliable online: <https://cloud.hzwtech.com/web/home>.
(56) Monkhorst, H. J.; Pack, J. D. Special points for Brillouin-zone integrations. *Phys. Rev. B* **1976**, *13* (12), 5188-5192.
(57) Liechtenstein, A. I.; Anisimov, V. I.; Zaanen, J. Density-functional theory and strong interactions: Orbital ordering in Mott-Hubbard insulators. *Phys. Rev. B* **1995**, *52* (8), R5467-R5470.
(58) Anisimov, V. I.; Zaanen, J.; Andersen, O. K. Band theory and Mott insulators: Hubbard U instead of Stoner I. *Phys. Rev. B* **1991**, *44* (3), 943-954.
(59) Kresse, G.; Furthmüller, J. Efficient iterative schemes for ab initio total-energy calculations using a plane-wave basis set. *Phys. Rev. B* **1996**, *54* (16), 11169-11186.
(60) Krukau, A. V.; Vydrov, O. A.; Izmaylov, A. F.; Scuseria, G. E. Influence of the exchange screening parameter on the performance of screened hybrid functionals. *J. Chem. Phys.* **2006**, *125 22*, 224106.
(61) Heyd, J.; Scuseria, G. E.; Ernzerhof, M. Hybrid functionals based on a screened Coulomb potential. *J. Chem. Phys.* **2003**, *118*, 8207-8215.
(62) Zaanen, J.; Sawatzky, G. A.; Allen, J. W. Band gaps and electronic structure of transition-metal compounds. *Phys. Rev. Lett.* **1985**, *55* (4), 418-421.

(63) Zesen Fu, M. H., Aolin Li, Haiming Duan, Junwei Liu and F. Ouyang. Strain-Controlled Topological Phase Transitions and Chern Number Reversal in Two-Dimensional Altermagnets. **2025**, *arXiv*:2507.22474.
(64) Dou, M.; Wang, X.; Tao, L. L. Anisotropic spin-polarized conductivity in collinear altermagnets. *Phys. Rev. B* **2025**, *111* (22), 224423.

**TOC**

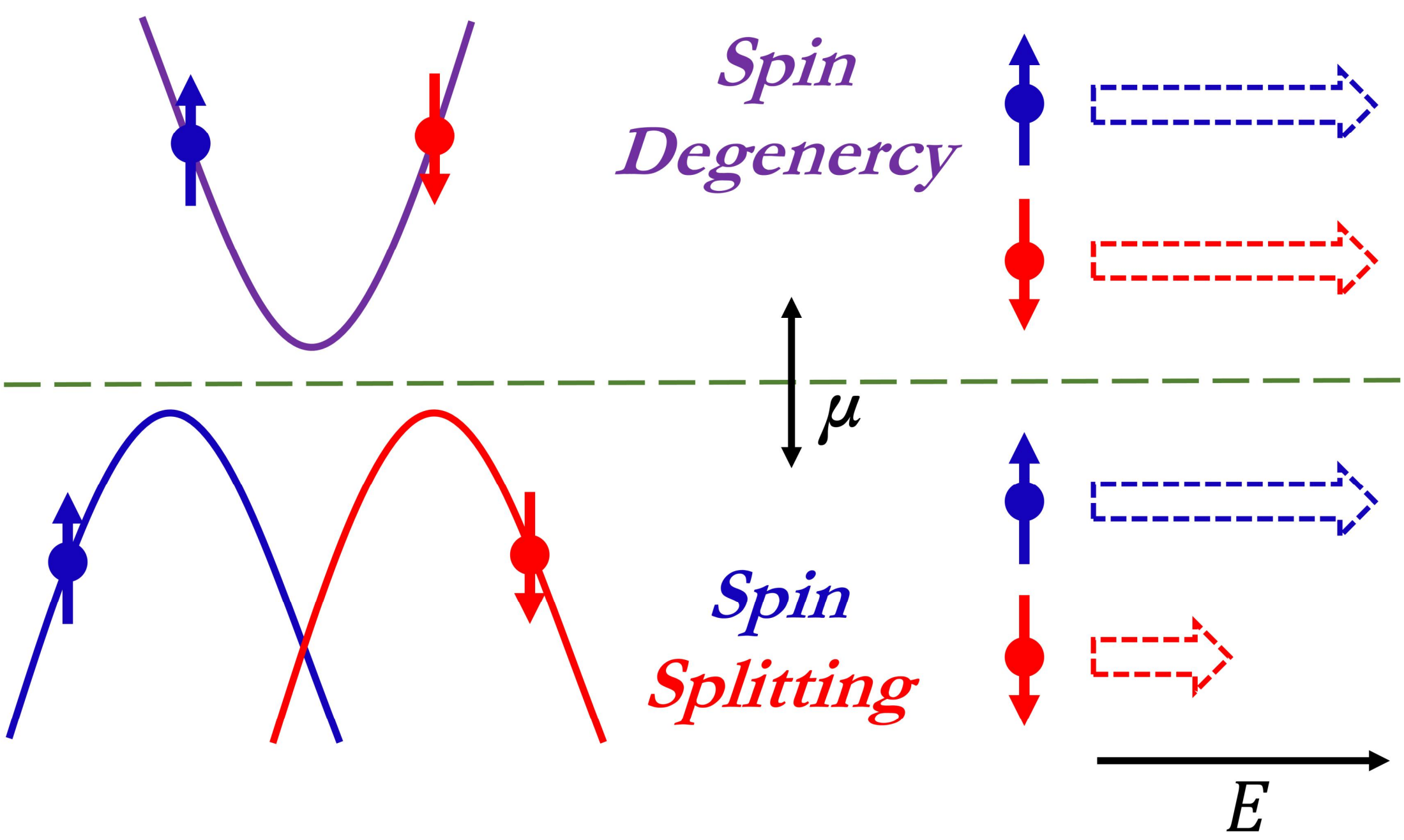